\documentclass[showpacs,twocolumn,prl,aps,color]{revtex4}
\usepackage{graphicx}
\begin{document}

\title{On the origin of the Fermi arc phenomena in the underdoped cuprates: signature of KT-type superconducting transition}
\author{Tao Li and Qiang Han}
\affiliation{Department of Physics, Renmin University of China,
Beijing 100872, P.R.China}

\date{\today}

\begin{abstract}
We study the effect of thermal phase fluctuation on the electron
spectral function $A(k,\omega)$ in a d-wave superconductor with
Monte Carlo simulation. The phase degree of freedom is modeled by a
XY-type model with build-in d-wave character. We find a ridge-like
structure emerges abruptly on the underlying Fermi surface in
$A(k,\omega=0)$ above the KT-transition temperature of the XY model.
Such a ridge-like structure, which shares the same characters with
the Fermi arc observed in the pseudogap phase of the underdoped
cuprates, is found to be caused by the vortex-like phase fluctuation
of the XY model.
\end{abstract}

\maketitle

The origin of the pseudogap phenomena in the underdoped cuprates is
among the most hotly debated issues in the high-T$_{c}$ physics.
Existing theories on the pseudogap falls into two categories: while
it is taken by many as the evidence for the existence of preformed
Cooper pairs and strong phase fluctuation in the normal
state\cite{Uemura,Kivelson,Orenstein,Xu}, some people believed that
it should be attributed to a yet another unidentified
order\cite{Lee,Chakravarty}. One puzzle about the pseudogap is its
momentum and temperature dependence. While early experiments
suggests that the pseudogap may inherit the d-wave structure of the
superconducting gap, more detailed ARPES measurement show that this
is not the case\cite{Ding,Kanigel,Damascelli}. Below $T^{*}$, the
pseudogap is first seen in a small momentum region around the
antinodal point $(\pi,0)$. The gaped region enlarges with decreasing
temperature until the whole Fermi surface(except the nodal points)
is gaped below $T_{c}$. The gradual development of the pseudogap on
the Fermi surface leaves the system with a finite segment of
ungapped Fermi surface which is called Fermi arc in the literature.
The length of the Fermi arc increases with temperature and is
reported to have a jump at $T_{c}$: the Fermi arc emerges abruptly
at $T_{c}$.

A metal with a unclosed Fermi surface is a strange animal in the zoo
of the Landau Fermi liquid theory. The Fermi surface, by its very
definition as an equal-energy contour at the chemical potential, is
always closed for any well defined quasiparticle dispersion. It is
proposed that the Fermi arc is just a half of a closed pocket-like
small Fermi surface whose other half is too weak to be detected by
ARPES\cite{Taillefer}. Such a scenario is widely adopted in the
second kind of theories for the pseudogap(related to an unidentified
order). In this paper, we try to understand the Fermi arc phenomena
in the framework of the phase fluctuation scenario. In this
scenario, the Fermi arc is not a genuine Fermi surface in the
momentum space across which a finite jump of the occupation number
occur, but a phase fluctuation induced pile up of the spectral
weight on the Fermi energy in a d-wave paired state.

The effect of the phase fluctuation on the electronic spectral
function have been studied by many
authors\cite{Millis,Eckl,Mayr,Berg}. However, most of these studies
are limited to the semiclassical treatment of the phase fluctuation.
At the semiclassical level, the phase fluctuation is modeled by a
uniform supercurrent with a phenomenological distribution of
velocity. The Doppler effect of the quasiparticle moving in such a
uniform supercurrent background then causes electronic spectral
function to shift in both the momentum and energy. More
specifically, when the center of mass momentum of the moving Cooper
pairs is $2q$, the electron pairing then happens between momentum
$q+k$ and $q-k$, and the excitation energy become
$E_{k}-\vec{v}_{k}\cdot \vec{q}$ to first order in $q$. The net
effect of the supercurrent on the electron spectral function is to
shift it by $q$ in momentum and $\vec{v}_{k}\cdot \vec{q}$ in
energy. Here $\vec{v}_{k}$ is velocity of the electron derived from
the bare band dispersion relation.

In a recent study\cite{Berg}, the author made the important
observation that in a d-wave superconductor the Doppler effect of
the fluctuating supercurrent will cause pile up of the spectral
weight exactly at the Fermi energy on the underlying Fermi surface,
no matter the detailed distribution of the fluctuating supercurrent.
The pile up of the spectral weight results in inverse square root
singularity at the Fermi energy in the electronic spectral function,
resembling a quasiparticle peak. The key point for this to happen is
that in a d-wave superconductor, the quasiparticle has linear
dispersion around the gap node. For such a system, the effect of the
momentum shift and the energy shift caused by the fluctuating
supercurrent cancels each other for $q$ in the nodal direction. For
$q$ in general direction, the transverse component of $q_{\perp}$
causes no energy shift in leading order and simply move the gap node
along the underlying Fermi surface transversely by a distance of
$q_{\perp}$.

The semiclassical approach, though intuitively attractive, has
limited value to give quantitative estimate of the phase fluctuation
effect. To simulate the actual temperature dependence of the phase
fluctuation effect, one must go beyond the long wave length limit
and take into account of the singular effect of the vortex
excitation on the quasiparticle motion. Even for the simulation of
the phase fluctuation involved in the far field of the vortex, the
semiclassical approach has the problem that it is unable to provide
a quantitative measure of the strength of the fluctuation.

In this paper, we report the results of the Monte Carlo simulation
of the effect of thermal phase fluctuation on the electronic
spectral function in d-wave superconductors. The phase fluctuation
is described in our work by a XY-type model with a build-in d-wave
character. The phase fluctuation then couples to the quasiparticles
of the d-wave superconductor via a standard pairing model with the
magnitude of the pairing potential $\Delta$ fixed. The simulation
shows clearly how the Fermi arc emerge abruptly above the KT
transition temperature of the XY model and then increase in strength
with temperature. These results indicate that the Fermi arc observed
in the underdoped cuprates is a phase fluctuation effect. The abrupt
emergence of the Fermi arc above $T_{c}$ suggests that the
superconducting transition of the underdoped cuprates are of KT
transition-type.

The phase degree of freedom is described by the following XY model
\begin{equation}
\mathrm{H}_{phase}=J\sum_{<\alpha,\beta>}\cos(\theta_{\alpha}-\theta_{\beta}),
\end{equation}
in which $\theta_{\alpha}$ is the phase variable defined on the
bonds of a square lattice on which the electrons reside. The center
of these bonds form another square lattice(see Fig.1) and the sum in
Eq.(1) is over neighboring sites of the latter square lattice. The
coupling constant $J$ is set to be positive so that a d-wave-like
phase structure is favored at low temperature in this model, namely
the phase variable on the neighboring bonds tends to have a phase
difference of $\pi$. Similar models are also used in other
works\cite{Eckl,Tesanovic}.

\begin{figure}[h!]
\includegraphics[width=6cm,angle=0]{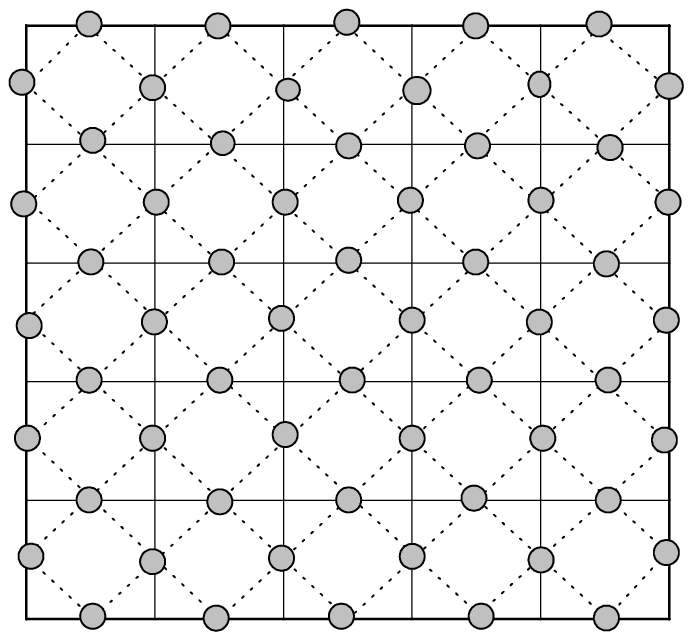}
\caption{The lattice on which model Eq.(1) is defined on. The
electrons reside on the square lattice made of the solid lines. The
phase variables resides on the bonds of this square lattice which
are indicated by the gray dots. These gray dots form another square
lattice of itself.} \label{fig1}
\end{figure}

In principle, the energetics of the phase degree of freedom should
be determined by the dynamics of the electron system itself. Here we
have treated them as separated degrees of freedom governed by a
coupling constant $J$. Except for being more simple to simulate,
such a model has the advantage that it takes into account the fact
that in the underdoped cuprates the spin pairing and phase coherence
are controlled by two separate energy scale\cite{Uemura,Lee}. Such a
separation of the energy scales is believed to the key for the
formation of the pseudogap phase in most scenarios.

The electron degree of freedom is described by the standard pairing
model of the form
$\mathrm{H}_{pair}=\mathrm{H}_{t}+\mathrm{H}_{\Delta}$, in which
$\mathrm{H}_{t}$ and $\mathrm{H}_{\Delta}$ are given by
\begin{eqnarray}
\mathrm{H}_{t}&=&-t\sum_{<i,j>,\sigma}(c^{\dagger}_{i,\sigma}c_{j,\sigma}+h.c.)\nonumber\\
&&-t'\sum_{<<i,j>>,\sigma}(c^{\dagger}_{i,\sigma}c_{j,\sigma}+h.c.)-\mu\nonumber
\end{eqnarray}
\begin{eqnarray}
\mathrm{H}_{\Delta}=\sum_{<i,j>}\left[\Delta
e^{i\theta_{\alpha}}(c^{\dagger}_{i,\uparrow}c^{\dagger}_{j,\downarrow}+c^{\dagger}_{j,\uparrow}c^{\dagger}_{i,\downarrow})+h.c.\right]
\end{eqnarray}
in which $t$ and $t'$ are the hopping integral of the electron on
the square lattice between neighboring and next neighboring sites.
The next neighboring hopping $t'$ is introduced to model the
curvature of the real Fermi surface. $\Delta$ denotes the magnitude
of the pairing potential and is kept as a constant in the whole
simulation. $\theta_{\alpha}$ is the phase angle on the bond between
site $i$ and $j$.

In the simulation, we first generate a thermal phase fluctuation
configuration from the distribution
$\rho=\exp(-\mathrm{H}_{phase}/k_{B}T)$. Then we diagonalize the
pairing Hamiltonian Eq.(2) with the generated phase fluctuation
configuration and calculated electronic spectral function. We then
average the electronic spectral function on different phase
fluctuation configurations.

For each individual phase fluctuation configuration, the
translational invariance of the pairing Hamiltonian is lost.
However, when the result is averaged over all possible phase
fluctuation configurations the translational invariance is
recovered. In the simulation, we have averaged the electron spectral
function over all configurations generated from a given one by space
translation at each step. After the average translational invariance
is recovered and electronic spectral function as a function of
momentum $A(k,\omega)$ is given by
\begin{equation}
A(k,\omega)=\sum_{n}|u^{n}_{k}|^{2}\delta(\omega-E^{n}),
\end{equation}
in which $u^{n}_{k}=\sum_{i}u^{n}_{i}e^{ikR_{i}}$. $u^{n}_{i}$
denotes the eigenvectors of the following B-dG equation with
eigenvalue $E^{n}$,
\begin{equation}
\left( {\begin{array}{cc}
H_{t,i,j}&H_{\Delta,i,j}\\
H^{*}_{\Delta,i,j}&-H_{t,i,j}\\
\end{array}}
\right)
\left( {\begin{array}{c}
u^{n}_{j}\\
v^{n}_{j}\\
\end{array}}
\right) = E^{n}\left( {\begin{array}{c}
u^{n}_{i}\\
v^{n}_{i}\\
\end{array}}
\right),
\end{equation}
in which
$H_{t,i,j}=-t\delta_{i,j+R_{\delta}}-t'\delta_{i,j+R_{\delta'}}-\mu\delta_{i,j}$
and $H_{\Delta,i,j}=\Delta
e^{i\theta_{\alpha}}\delta_{i,j+R_{\delta}}$ are the matrix element
of $\mathrm{H}_{t}$ and $\mathrm{H}_{\Delta}$. Here sum over
repeated indexes is assumed and $R_{\delta}$ and $R_{\delta'}$
denotes the neighboring and next neighboring vector on the square
lattice. When $A(k,\omega)$ is calculated for a given phase
fluctuation configuration, we do the average over the configurations
by the standard Monte Carlo procedure
\begin{equation}
\langle A(k,\omega)\rangle=\frac{\sum
e^{-\mathrm{H}_{phase}/k_{B}T}A(k,\omega)}{\sum
e^{-\mathrm{H}_{phase}/k_{B}T}}
\end{equation}

\begin{figure}[h!]
\includegraphics[width=7cm,angle=0]{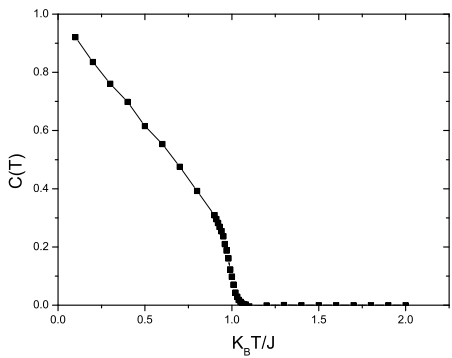}
\caption{The correlation(defined in the text) between the phase
variables on the most far apart bonds of a $64\times64$ lattice as a
function of temperature. The abrupt drop of this correlation around
$k_{B}T/J\sim1$ indicates the KT transition of that finite lattice.}
\label{fig2}
\end{figure}
\begin{figure*}
\includegraphics[width=16cm,angle=0]{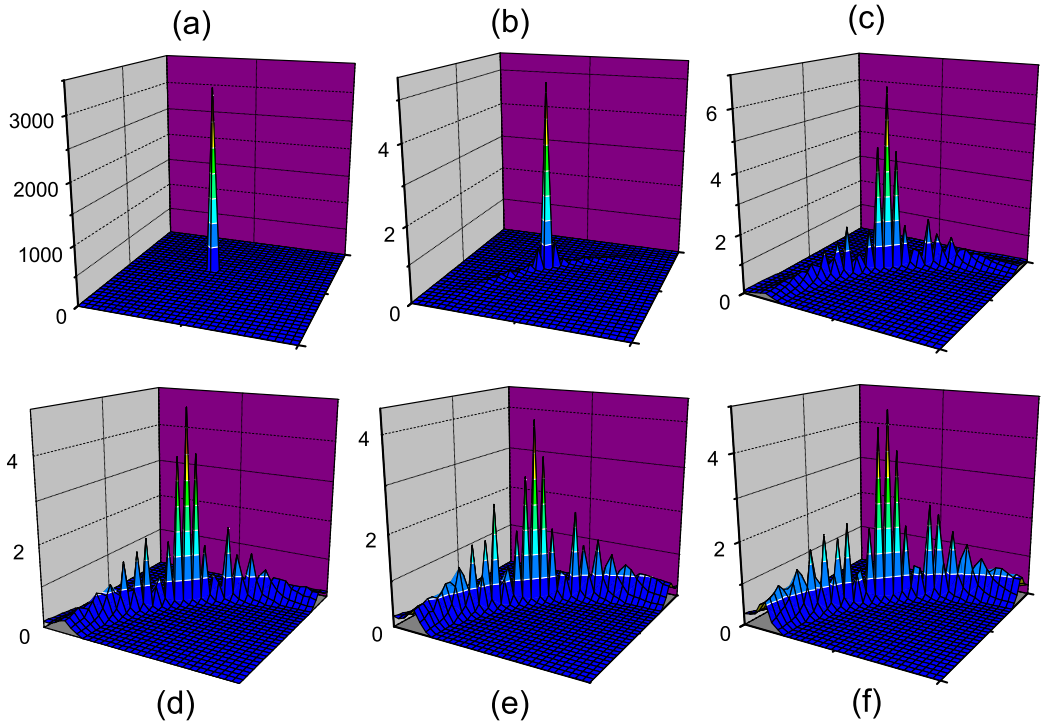}
\caption{The temperature dependence of the electron spectral
function at the Fermi energy $A(k,\omega=0)$. (a)$k_{B}T=0.5J$,
(b)$k_{B}T=J$, (c)$k_{B}T=1.1J$, (d)$k_{B}T=1.2J$,
(e)$k_{B}T=1.3J$,(f)$k_{B}T=1.4J$. Shown here are the results in the
first quadrant of the Brillouin zone only.} \label{fig5}
\end{figure*}

The diagonalization of the B-dG Hamiltonian is the most time
consuming part of our simulation. At the same time, to resolve the
Fermi surface on the finite momentum mesh, we have to study lattice
of large enough size. To study relatively larger lattice so that the
Fermi surface can be resolved, we adopt the following cutoff
strategy. First, we Fourier transform(by FFT algorithm) the B-dG
Hamiltonian into momentum space. The transformed Hamiltonian reads
\begin{equation}
\left( {\begin{array}{cc}
\xi_{k}\delta_{k,k'}&\Delta_{k,k'}\\
\Delta_{k,k'}&-\xi_{k}\delta_{k,k'}\\
\end{array}}
\right) \left( {\begin{array}{c}
u^{n}_{k'}\\
v^{n}_{k'}\\
\end{array}}
\right) = E^{n}\left( {\begin{array}{c}
u^{n}_{k}\\
v^{n}_{k}\\
\end{array}}
\right),
\end{equation}
in which $\xi_{k}=-2t(\cos k_{x}+\cos k_{y})-4t'\cos k_{x}\cos
k_{y}-\mu$, $\Delta_{k,k'}=1/N\sum_{<i,j>}\Delta
e^{i\theta_{\alpha}}e^{ikR_{i}-ik'R_{j}}$. To simulate the Fermi arc
phenomena, we only need to calculate the electron spectral function
right on the Fermi energy, $A(k,\omega=0)$. Obviously, those
momentums for which $|\xi_{k}|\gg\Delta$ are not expected to
contribute significantly to $A(k,\omega=0)$. For this reason, we set
a cutoff energy $E_{c}$ and neglect all momentums for which
$|\xi_{k}|>E_{c}$. We then diagonalize the truncated B-dG
Hamiltonian below the cutoff energy and calculate the electron
spectral function on the Fermi energy. The convergence of such a
cutoff strategy can be easily checked by varying the cutoff energy
$E_{c}$. In real simulation, we find $E_{c}=5\Delta$ is large
enough.

We now present the results of our simulation. The simulation is done
on a $64\times64$ lattice. We have used $t'/t=-0.3$ and
$\Delta/t=0.1$ to model the real system. The coupling strength of
the phase variables, $J$, is set as the unit for temperature. The
strength of the phase fluctuation can be read out from the
correlation function of the phase variables. In Figure 2, we plot
the correlation $C(T)=\langle\cos
(\theta_{\alpha}-\theta_{\alpha'})\rangle$ between the phase
variables on the most far apart bonds of the $64\times64$ lattice as
a function of temperature. For infinite system, such correlation is
zero at finite temperature as required by the Wagner-Mermin theorem.
The abrupt drop of this correlation signifies the KT transition of
the finite system. For our $64\times64$ lattice, we find the abrupt
drop of the phase correlation occurs around $k_{B}T/J\sim1$,
slightly above the KT transition temperature for the infinite system
($k_{B}T\sim 0.8923J$)\cite{Olsson,Hasenbusch}.

To achieve statistical independence of the samples generated from
the Monte Carlo procedure, we have used $10^{5}$ local update of the
phase variables to generate every new phase fluctuation
configuration. The first $10^{4}$ configurations are discarded for
thermalization. The electron spectral function is averaged over 1000
independent phase fluctuation configurations. Figure 3 shows the
$A(k,\omega=0)$ calculated at six temperatures. Among the six
temperatures, one($k_{B}T=0.5J$) is below the KT transition point
and the other five are above it($k_{B}T=J,1.1J,1.2J,1.3J,1.4J$).

Below the KT transition temperature($T_{KT}$), the phase fluctuation
effect is almost totally quenched and the electron spectral function
at the Fermi energy is composed of four sharp peaks at the four
nodal points, as in an ideal d-wave superconductor. Right above the
KT transition temperature, a ridge-like structure in $A(k,\omega=0)$
emerges on the underlying Fermi surface\cite{oscillation}. The
intensity of this ridge-like structure increases rapidly in a small
temperature region between $k_{B}T=J$ and $k_{B}T=1.1J$. For
$k_{B}T>1.1J$, the increase of electron spectral function on the
Fermi energy becomes much slower. Such temperature dependence of the
electron spectral function at the Fermi energy is in close
resemblance to that observed in experiments\cite{Kanigel}. It should
be noted that the ridge-like structure emerges simultaneously on the
whole Fermi surface, though it is clear anisotropic on the Fermi
surface. However, considering the finite resolution of the ARPES
measurement, it is still reasonable to define a finite segment of
the Fermi surface as the Fermi arc for each temperature by setting
some threshold for the spectral weight. The arc length defined in
such a way clearly increases with temperature.

To quantify the temperature dependence of the electron spectral
function in a more objective way, we plot in Figure 4 the
integration of the electron spectral function at Fermi energy in the
Brillouin zone $W(\omega=0,T)=\int dk A(k,\omega=0)$ as a function
of temperature. The quench of the phase fluctuation effect below
$T_{KT}$ and the abrupt emergence of the ridge-like structure above
it can be clearly seen in this figure.

We thus conclude that the main features of the experimental
observations concerning the Fermi arc phenomena can be accounted for
by our phase fluctuation model. The quench of the phase fluctuation
effect below $T_{KT}$ and the its abrupt emergence above $T_{KT}$
indicates that Fermi arc phenomena is controlled by the same
mechanism that induce the KT transition, namely the vortex
excitation. Indeed, according to the study of the XY model, the
vortex density enjoys a proliferation just above the KT transition
point. With these understanding, it is tempting to argue that the
Fermi arc behavior observed in the high-Tc cuprates provide strong
evidence that the superconducting transition in the underdoped
cuprates are of KT transition-type.

\begin{figure}[h!]
\includegraphics[width=8cm,angle=0]{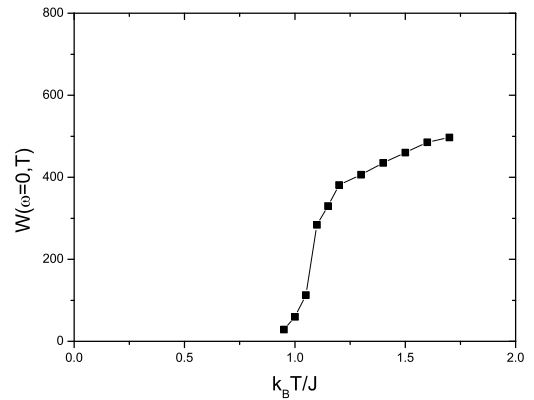}
\caption{The integrated electron spectral function over momentum at
the Fermi energy $W(\omega=0,T)=\int dk A(k,\omega=0)$ as a function
of temperature.}\label{fig4}
\end{figure}

This work is supported by NSFC Grant No. 10774187, National Basic
Research Program of China No.2007CB925001 and Beijing Talents
Program.

\bigskip


\begin{thebibliography}{99}
\bibitem{oscillation}
The strong oscillation of the electron spectral function on the
Fermi surface in Fig.3 is not due to the statistical error. The
oscillaition has a fixed pattern and is caused by the finite lattice
size effect. On a finite lattice, the momentum mesh in the Brillouin
zone are not dense enough to resolve all the momentum on the Fermi
surface and as a result some Fermi surface momentums are missed(or
approximated by nearby momentums).
\end{thebibliography}
\end{document}